\documentclass[aps,prb,twocolumn,showpacs,floatfix]{revtex4}

\newcommand{\beq}{\begin{equation}}
\newcommand{\eeq}{\end{equation}}
\newcommand{\up}{\uparrow}
\newcommand{\down}{\downarrow}
\usepackage{graphicx}
\usepackage{amsmath}
\graphicspath{{.}{./EPS/}}

\begin{document}

\title{Rashba spin precession in quantum Hall edge channels}

\author{Marco G.~Pala}
\affiliation{Dipartimento di Ingegneria dell'Informazione,
Universit\`a degli Studi di Pisa,
via Caruso, I-56122 Pisa, Italy}
\affiliation{Institut f\"ur Theoretische Festk\"orperphysik,
Universit\"at Karlsruhe, D-76128 Karlsruhe, Germany}
\author{Michele Governale}
\affiliation{NEST-INFM \& 
Scuola Normale Superiore, piazza dei Cavalieri 7, I-56126 Pisa, Italy}
\author{Ulrich Z\"ulicke}
\affiliation{Institute of Fundamental Sciences, Massey 
University, Private Bag 11~222, Palmerston North, New Zealand}
\author{Giuseppe Iannaccone}
\affiliation{Dipartimento di Ingegneria dell'Informazione,
Universit\`a degli Studi di Pisa,
via Caruso, I-56122 Pisa, Italy}
\affiliation{IEIIT-Consiglio Nazionale delle Ricerche, via Caruso,
I-56122 Pisa, Italy}
\date{\today}

\begin{abstract}
Quasi--one dimensional edge channels are formed at the boundary of a
two-dimensional electron system subject to a strong perpendicular
magnetic field. We consider the effect of Rashba spin--orbit coupling,
induced by structural inversion asymmetry, on their electronic and
transport properties. Both our analytical and numerical results show
that spin--split quantum--Hall edge channels exhibit properties
analogous to that of Rashba--split quantum wires. Suppressed
backscattering and a long spin life time render these edge channels an
ideal system for observing voltage--controlled spin precession. Based
on the latter, we propose a magnet--less spin--dependent electron
interferometer.

\end{abstract}

\pacs{85.75.-d,73.43.-f,73.63.-b}

\maketitle

\section{Introduction}
\label{s1}

Spin--dependent transport in semiconductors has attracted a lot of
interest recently, due to intriguing new physics phenomena that are
observed experimentally or predicted theoretically.\cite{lossbook,zutic}
Some of these may form the basis of future device applications within
the spintronics paradigm, where information is coded and transferred
using the spin degree of freedom instead of charge.\cite{wolf} External
magnetic fields and magnetic contacts provide a possible means to control
the spin of charge carriers.\cite{prinzreview} 
Spin control via spin--orbit (SO) coupling, which is interesting from a
fundamental--physics viewpoint\cite{datta} and possibly useful for
device application, \cite{devices} has gained prominence recently
as an intriguing alternative to the use of magnetic systems.
In particular, the Rashba--type SO
coupling\cite{rashba,lommer,andrada1} which arises from structural
inversion asymmetry in semiconductor heterostructures is of particular
interest to spintronics research, as its strength can be tuned by
external gate voltages.\cite{nitta,schaepers,grundler,yamada}

At the same time as enabling novel spin--dependent transport effects,
SO coupling is also responsible for spin--relaxation phenomena that limit
the operation of spintronics devices. For mesoscopic electron transport
in semiconductors, the Dyakonov--Perel mechanism\cite{dyakonov} is the
dominant source of spin relaxation. It is due to elastic scattering which
randomizes the orientation of momentum and, via SO coupling, the spin
orientation. This mechanism limits experimental observation of coherent
spin--dependent transport phenomena, such as spin precession in magnetic
fields or due to the Rashba effect.\cite{datta}

Here we consider a system well--suited to the study of spin--dependent
transport effects, due to its relatively weak spin relaxation: a
two--dimensional (2D) electron system in the integer quantum Hall (QH)
regime.\cite{smgrev} The latter is realized when a perpendicular magnetic
field $B$ is applied such that the filling factor $\nu=2\pi l_B^2
n_{\text{2D}}$ assumes integer values (here $l_B=\sqrt{\hbar/|e B|}$ is
the magnetic length, and $n_{\text{2D}}$ the electronic sheet density).
Then a bulk incompressibility occurs in the 2D electron system and, for a
sufficiently steep confining potential at the sample boundaries,
transport is possible only via chiral quasi--one dimensional edge
channels.\cite{halperin,mac,stripescaveat} The spatial separation of
right--moving and left--moving edge channels by the incompressible bulk
prevents backscattering. Furthermore, long equilibration lengths between
opposite--spin QH edge states (of the order of $100\,\mu$m) have been
observed in GaAs--based samples.\cite{gerhard,axel} Spin flips induced by
impurity scattering in the presence of SO coupling were
found\cite{gerhard,khaetskii} to be the dominant mechanism for spin
relaxation in the QH regime. Hence, the typically stronger SO coupling in
InAs--based 2D heterostructures should reduce spin life times for QH edge
channels realized in such samples. Indeed, for identical quantum--well
parameters (such as width, sheet and donor densities) and edge--channel
profiles, the ratio of spin--relaxation lengths $l_{\text{sf}}$ in the
InAs and GaAs materials systems can be approximated, in the high--field
limit, by\cite{khaetskii,polyakov}
\begin{equation}
\frac{l_{\text{sf}}^{\text{InAs}}}{l_{\text{sf}}^{\text{GaAs}}}\approx
\left|\frac{g_{\text{InAs}}}{g_{\text{GaAs}}}\right|\left(\frac{L_{\text
{so}}^{\text{InAs}}}{L_{\text{so}}^{\text{GaAs}}}\right)^2 \sim 0.1
\mbox{ typically} \quad .
\end{equation}
We have denoted the gyromagnetic ratio by $g$, and $L_{\text{so}}$ is the
spin--precession length due to the strongest spin--orbit coupling
present in the respective materials (Rashba--type for InAs,
Dresselhaus--type\cite{2Ddressel} for GaAs\cite{khaetskii}). From our
rough estimate, we would expect the spin--flip length $l_{\text{sf}}$ in
InAs--based QH edge channels to exceed tens of microns, which is much
larger than the spin--{\em precession\/} length $L_{\text{so}}$ that
determines, e.g., the gate length in a spin--controlled field--effect
transistor.\cite{datta,epl} This motivates our study of spin--dependent
transport in Rashba--split QH edge channels.

Previous studies of magnetotransport in the presence of Rashba spin
splitting have focused on beating patterns in the Shubnikov--de~Haas
oscillations,\cite{tarasenko,vasilopoulos} the Hall resistance,\cite
{vasilopoulos} and quasi--one dimensional point--contact
conductances.\cite{wang} The interplay between spin--orbit coupling and
cyclotron motion was discussed recently\cite{usaj} in the
weak--magnetic--field regime with particular emphasis on using magnetic
focusing to separate electrons according to their spin
state.\cite{rokhinson}

In this paper we present a theory for QH edge states with Rashba
SO coupling in the strong field regime where the Rashba spin splitting is
much smaller than the cyclotron energy. We give an analytical
approximation for the Landau--level dispersions, both for the case when
the Zeeman term is negligible and when the Zeeman splitting is comparable
to the Rashba one. Our analytical results show that QH edge states in the
presence of Rashba SO coupling behave, as far as spin precession is
concerned, in a very similar way to Rashba--split quantum wires.\cite
{mireles,rashbawire,wang} Furthermore, we study spin transport in edge
channels by means of the numerical recursive--Green's--function technique
without making any of the approximations that were necessary to obtain
analytical results. The numerical transport calculations allow us to
test the validity of our analytical Landau--level description when used
to describe linear transport. 

The paper is organized as follows. We introduce the theoretical
description in Sec.~\ref{s2} and give analytical results on Landau--level
dispersions. In Sec.~\ref{s4} we present numerical results on
spin--dependent transport and test the validity of approximations made in
Sec.~\ref{s2}. Sec.~\ref{s5} is devoted to a discussion of an
interferometer setup, suitable for observing interference effects due to
spin precession. Conclusions are given in Sec.~\ref{conclusions}.

\section{Theoretical description and analytical results}
\label{s2}
In this section, we introduce the model Hamiltonian for our system of
interest. Analytical results are presented, within certain
approximations, for edge--channel energy dispersions and wave functions
with Rashba SO coupling. 

We study a two--dimensional electron system in the $xy$ plane, subject to
a homogeneous perpendicular magnetic field ${\bf B}=B {\bf \hat{z}}$, and
confined laterally (in $y$ direction) by the boundary potential $V(y)$.
Translational invariance in $x$ direction suggests the use of the Landau
gauge with vector potential ${\bf A}=-B\,y\, {\bf \hat{x}}$. Furthermore,
we assume that the electrons are subject to a SO coupling of the Rashba
type,\cite{rashba} and neglect the SO coupling arising from bulk
inversion asymmetry.\cite{dresselhaus,2Ddressel}. This is reasonable as a
first approximation to describe realistic InAs quantum--well
systems.\cite{koga,ganichev} The Hamiltonian of the system is then given
by $H=H_0+H_{\text{R}}+H_{Z}$, with
\begin{subequations}
\begin{eqnarray}
\label{h0}
H_0&=&\frac{1}{2m^*} \left[ (p_x+ e By)^2+p_y^2 \right]+V(y) \;,
\label{rh}\\
H_{\text{R}}&=&\frac{\alpha_{\rm R}}{\hbar} 
[\sigma_x p_y - \sigma_y (p_x+eBy)] \;,\\ \label{Zeemanham}
H_{\text{Z}}&=&\frac{g}{2} \mu_B B \; \sigma_z=\frac{\nu_{\text{Z}}}{2}
\sigma_z\;,
\end{eqnarray}
\end{subequations}
where $m^*$ is the effective mass, $e$ ($<0$) the electron charge,
$\vec{p}$ the canonical momentum, $\alpha_{\text{R}}$ measures the
strength of Rashba SO coupling,
and $\mu_{\text{B}}=|e|\hbar/2m_e$ denotes the Bohr magneton. 
In the following, it will be useful to express
the SO coupling strength in terms of a length scale, $l_{\text{R}}=
\hbar^2/(m^\ast \alpha_{\text{R}})$, which is related to the
spin--precession length\cite{datta,epl} $L_{\text{so}}$ mentioned in the
previous section via $l_{\text{R}}=L_{\text{so}}/\pi$. Our study focuses
on the high--field regime where $l_B < l_{\text{R}}$.

\subsection{Results in the absence of Zeeman splitting}

We start by discussing the case of vanishing Zeeman splitting. The
validity of this approximation is discussed in the next subsection, where
the effect of the Zeeman splitting on the edge states is studied at the
level of perturbation theory. We furthermore neglect the term 
$(\alpha_{\rm R}/\hbar) \sigma_x p_y $ from the Rashba Hamiltonian.
This approximation, which we call {\em longitudinal SO
approximation},\cite{mireles} turns out to be valid, in the high--field
regime, when the transverse width of the QH states 
is smaller than the SO length $l_{\text{R}}$ 
and the transport becomes quasi--one dimensional 
[note that the transverse Rashba term
$(\alpha_{\rm R}/\hbar)\sigma_x p_y$ becomes important, e.g., at
Landau--level crossings\cite{usaj,falko}]. 
The Hamiltonian can now be written as
\begin{equation}
\label{happrox}
\tilde{H}=\frac{p_y^2}{2m^*}+\frac{m^*\omega_{\text{c}}^2}{2}
\left(\hat{Y}+y-\frac{l_B^2}{l_{\text{R}}}\sigma_y \right)^2+V(y),
\end{equation}
where $\hat{Y}=(p_x/\hbar) l_B^2 \text{sgn}(eB)$ is the operator of
guiding--center coordinate for cyclotron motion, and $\omega_{\text{c}}=
|e B|/m^{*}$ the cyclotron frequency. In writing Eq.~(\ref{happrox}), we
have neglected a constant energy shift of order $\hbar^2/(m^\ast
l_{\text{ R}}^2)$, which is small compared to the cyclotron gap. For the
eigenfunctions of Hamiltonian~(\ref{happrox}), we make the
\textit{Ansatz}
\begin{equation}
\label{ansatz}
\psi_{n,Y,\sigma}(x,y)=e^{iY x/l_{B}^2} 
\phi_{n,Y,\sigma}(y)|\sigma_y\rangle  \;, 
 \end{equation}
where $|\sigma_y\rangle$ is the eigenspinor of the Pauli matrix
$\sigma_y$ with eigenvalue $\sigma=\pm 1$. Substituting the
\textit{Ansatz} Eq.~(\ref{ansatz}) in the Schr\"odinger equation, we find
the eigenenergies
\begin{equation}
\label{landaulevels}
E_n(Y)=E_n^{(0)}\left( Y-\sigma \frac{l_B^2}{l_\text{R}} \right) \quad .
\end{equation}
Here $E_n^{(0)}(Y)$ are the Landau--level dispersions in the absence of
Rashba SO coupling. The transverse eigenfunctions are given by 
\begin{equation}
\label{transwf}
\phi_{n,Y,\sigma}(y)= \phi^{(0)}_{n,Y-\sigma\frac{l_B^2}{l_\text{R}}}(y),
\end{equation}
where $\phi^{(0)}_{n,Y}(y)$ are the corresponding transverse
eigenfunctions without Rashba SO coupling. 

\begin{figure}[t]
\begin{center}
\includegraphics[width=3.1in]{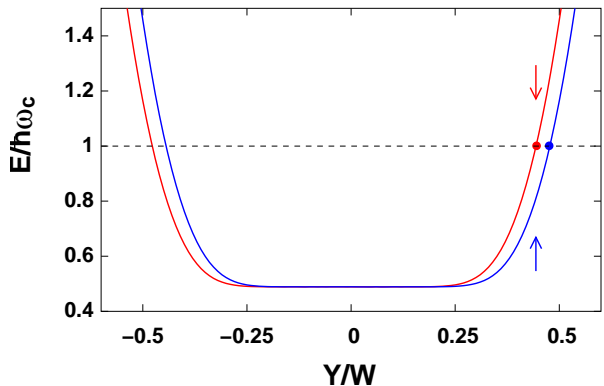}
\end{center}
\caption{(Color online)
Dispersion of the lowest Landau level, calculated within the longitudinal
SO approximation, for a hard--wall confining potential of width $W$.  
The labels $\uparrow,\downarrow$ refer to eigenvalues of $\sigma_y$. 
The parameters used are: $L_{\text{so}}/l_B=21.3$, and $W/l_B=8.9$.}
\label{fig1}
\end{figure}
At this point a few comments are necessary. Within the longitudinal SO
approximation, the effect of the Rashba SO coupling is to shift Landau
levels with different spin quantum number $\sigma$ in the guiding--center
quantum number $Y$. An example of a Landau--level dispersion is shown in
Fig.~\ref{fig1}. In this limit, a global spin--quantization axis exists,
which is perpendicular to the system boundary and in the plane of the 2D
electron system. This is analogous to what happens in quantum wires in
the weak--SO--coupling regime.\cite{rashbawire} As these properties are
those on which the design of a spin--controlled field--effect transistor
(SpinFET) relies,\cite{datta} we can conclude that a prototype of the
SpinFET could be implemented using QH edge channels. Such a system would
realize the ideal situation of a highly one dimensional transport
regime\cite{wang} and slow spin relaxation from elastic scattering.

\begin{figure}[t]
\begin{center}
\includegraphics[width=3.1in]{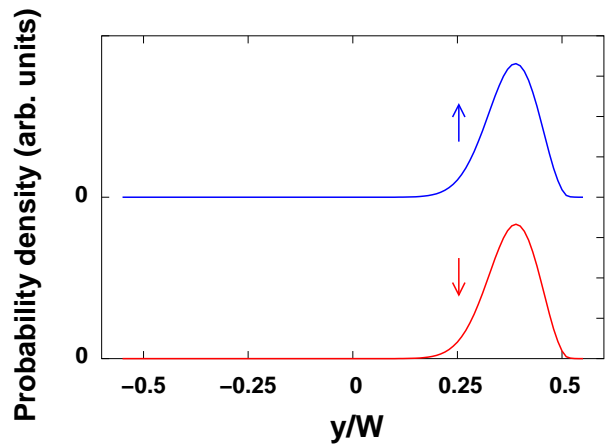}
\end{center}
\caption{(Color online)
Transverse probability--density profile for spin up (top) and spin down
(bottom) right-moving edge states at a fixed energy $E=\hbar\omega_{\text
{c}}$. The transverse probability densities, shown here, correspond to
the states marked by a dot in Fig.~\ref{fig1}. The spin labels $\uparrow,
\downarrow$ refer to eigenvalues of $\sigma_y$. Parameters are the same
as in Fig.~\ref{fig1}.}
\label{fig2}
\end{figure}

The quantum numbers $Y_{n,\sigma}$ of the guiding--center coordinate in
$y$ direction for edge states at a fixed energy $E$ are given by  
\beq
\label{yns}
Y_{n,\sigma}=
Y_{n}^{(0)}+\sigma l_{B}^2/
l_{\text{R}},
\eeq
where $Y_{n}^{(0)}$ satisfies $E_n^{(0)}(Y_{n}^{(0)})=E$. The
guiding--center separation for states at fixed energy turns out to not
depend on this energy or the Landau--level index $n$; we find
\beq
\label{delta}
\Delta Y =2l_{B}^2/l_{\rm R} \;.
\eeq 
It is important to notice that the density profile in confinement
direction for wave functions~(\ref{transwf}) corresponding to spin--split
edge states [i.e., with guiding centers $Y_{n,\sigma}$ given in
Eq.~(\ref{yns})] is the same, and is simply $|\phi^{(0)}_{Y_{n}^{(0)}}(y)|^2$. 
This last remark means that although edge states with different spin
are shifted in their guiding--center {\em quantum number}, they
\textit{are not separated spatially} (see Fig.~\ref{fig2}). 
 
For small $l_{B}/l_{\text{R}}$, we can expand the Landau level dispersion
Eq.~(\ref{landaulevels}) to first order around $Y$, obtaining
\begin{equation}
\label{ll1}
E_n(Y)
\approx
E_n^{(0)}(Y)-\sigma\frac{l_B^2}{l_\text{R}}\frac{\partial E_n^{(0)}(Y)}
{\partial Y}.  
 \end{equation}
From Eq.~(\ref{landaulevels}), and even more from Eq.~(\ref{ll1}), it is
apparent that bulk Landau levels, which are non dispersive, are not
affected to first order in $l_{B}/l_{\text{R}}$ by Rashba SO coupling.
This is in agreement with the exact solution for bulk Landau levels given
in  Ref.~\onlinecite{rashba}, where the first non vanishing correction to
eigenenergies is quadratic in  $l_{B}/l_{\text{R}}$.

\subsection{Effect of finite Zeeman splitting}
\label{s3}
In the previous subsection, we neglected the Zeeman effect. We now
discuss how a finite but small Zeeman splitting affects spin precession
in QH edge channels. 

The value of spin splitting due to the Zeeman effect can be expressed as
\beq
\label{zee}
\nu_{\rm Z}=\frac{gm^*}{2m_e} \hbar \omega_c \quad ,
\eeq
whereas the Rashba spin spitting for an edge state with guiding center
$Y$ is given by
\beq
\label{rashba}
\nu_{\rm R}(Y)={2}\,\frac{l_B^2}{l_\text{R}}\,
\frac{\partial E_n^{(0)}(Y)}{\partial Y}= \frac{2\hbar}{l_{\rm R}}\,
 v_n^{(0)}(Y) \quad .
\eeq
Here $v_n^{(0)}=(l_{\text{B}}^2/\hbar)\, \partial E_n^{(0)}/ \partial Y$
is the group velocity of edge states on the $n^\text{th}$ unperturbed
Landau level. For a boundary confining potential that is rising sharply
on the scale of the magnetic length, it can be estimated as $v_n^{(0)}
\approx\omega_{\text{c}} l_B$, yielding
\beq
\label{comparison}
\left|\frac{\nu_{\rm Z}}{\nu_{\rm R}}\right|= \frac{g}{4}\,\frac{m^\ast}
{m_{\text{e}}}\,\frac{l_{\text{R}}}{l_B} = \frac{0.7  \; g \;  
\sqrt{B[T]}}{\alpha_{\text{R}}[10^{-12}\,\mbox{eV m}]} \quad .
\eeq
For typical values in InAs heterostructures,\cite{thomasnew} the Zeeman
splitting becomes comparable to the Rashba spin splitting at a magnetic
field of $\sim 8$~Tesla. Note, however, that edge velocities at soft
sample--boundary potentials can be an order of magnitude smaller than the
estimate used above.

In the situation when the Zeeman and the Rashba splitting are comparable 
and both  are much smaller 
than $\hbar\omega_{\text{c}}$, we can perform a perturbative 
calculation, finding for the Landau-level dispersions\cite{khaetskii}:
 \beq
\label{ezeeman}
E_{n}^\pm(Y)=
E^{(0)}_n(Y) \mp \frac{1}{2} \sqrt{\nu_{\rm R}(Y)^2+\nu_{\rm Z}^2} \;.
\eeq
After performing a perturbative calculation on a spin--degenerate
subspace, we find that the orbital part of the eigenfunctions is
unchanged, while the eigenspinors read
\begin{eqnarray}
\nonumber
\chi^{+}(Y) &=& 
\left( \begin{array}{c} \sin [\theta(Y)/2]\\ i \cos [\theta(Y)/2]
\end{array} \right) ,\quad\\
\label{szeeman}
\chi^{-}(Y) &=& 
\left( \begin{array}{c} \cos [\theta(Y)/2)]\\ -i \sin [\theta(Y)/2]
\end{array} \right),
\end{eqnarray}
with $\tan[\theta(Y)]= \nu_{\text{Z}}/\nu_{\text{R}}(Y)$.
If we set $\nu_{\text{Z}}=0$ in Eqs. ~(\ref{ezeeman},\ref{szeeman}) we
find the Landau level dispersions given in Eq.~(\ref{ll1}), and the
eigenspinors become $|\sigma_y\rangle$. In the presence of the Zeeman
term, the eigenspinor quantization axis depends on $Y$. In particular, it
does not lie anymore in the $xy$ plane, but it sticks out of it; the
in--plane component remaining still perpendicular to the boundary (i.e.,
parallel to $y$). The new eigenspinors Eq.~(\ref{szeeman}) are parallel
to the effective magnetic field $\mathbf{B}_{\rm eff}(Y)=\mathbf{B}+
\mathbf{B}_{\text{R}}(Y)$, where $\mathbf{B}_{\text{R}}$ is the effective
Rashba field (the Rashba term can be viewed as a Zeeman term with a
momentum dependent magnetic field), which in our case is $\mathbf{B}_{\rm
R}(Y)=- (\nu(Y) / g \mu_{\text{B}}) \hat{y}$.

\section{Spin--dependent transport: Numerical results}
\label{s4}

Our results obtained in the previous section showed that, within the
longitudinal SO approximation and for vanishing Zeeman splitting, QH
edge--channel eigenspinors are polarized in the direction perpendicular
to the sample boundary and in the plane of the QH system. In that
situation, spin precession will occur for electrons injected
with spins parallel to the edge, e.g., by a magnetic contact. This is
entirely analogous to the operational principle of a SpinFET.\cite{datta}
To test the validity of the underlying approximations made to obtain our
analytical results, we have studied spin--dependent edge--channel
transport, in the presence of  Rashba SO coupling and Zeeman splitting,
numerically without making any of the approximations of the previous
section.

We compute total and spin--polarized linear conductances of a
finite--size QH system within the framework of the Landauer--B\"uttiker
theory,\cite{landauer} assuming the zero--temperature limit. A
tight--binding model is adopted to describe the Hamiltonian of the
system,\cite{mireles} and we use a recursive method to obtain the total
Green's function of the system.\cite{baranger,ferry,frustaglia}
Projecting the Green's function on asymptotic waves in the leads and on
spin--up and spin--down eigenspinors, transmission and reflection
coefficients are immediately obtained.\cite{fisherlee,ferry} The
conductance is expressed by the Landauer--B\"uttiker formula
\beq
G=\frac{e^2}{h}\sum_{nn'}\sum_{\sigma\sigma'} |t_{n'n}^{\sigma'\sigma}|^2
\;,
\eeq
where the sum runs over all incoming and outgoing channels, and
$t_{n'n}^{\sigma'\sigma}$ is the transmission amplitude 
from mode $n$ with spin $\sigma$
to mode $n'$ having spin $\sigma'$. We assume that the system is attached to
external leads with the same homogeneous magnetic field as present in the
sample, but without SO coupling. In the external leads, we choose $z$ as 
spin quantization axis and, in this Section, ``up'' and ``down'' always
refers to eigenspinors of $\sigma_z$. Together with the total
conductance, we calculate $G_{\sigma \up}$, i.e., the conductance
obtained by injecting a spin-up current and detecting a spin-$\sigma$
current at the output contact. 

\begin{figure}[t]
\begin{center}
\includegraphics[width=3.1in]{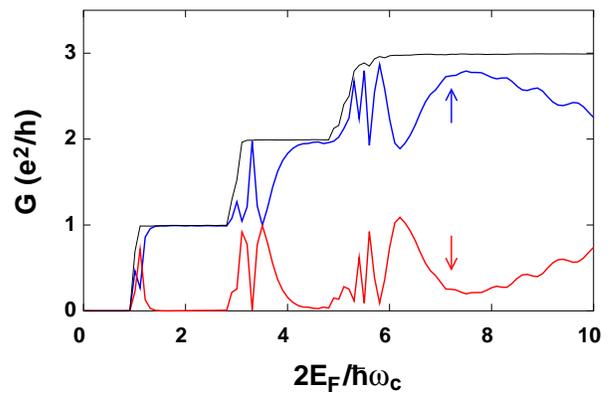}
\end{center}
\caption{(Color online) Spin--polarized conductances as a function of
the inverse magnetic field $1/\hbar\omega_{\text{c}}$ for fixed Fermi
energy $E_{\text{F}}$. The blue line represents $G_{\up \up}$, the red
line  $G_{\down \up}$, and the  black line $G_{\up \up}+G_{\down \up}$.
The Zeeman coupling is set to zero. The other parameters are: 
$L/\lambda_{\text{F}}=40$, $W/\lambda_{\text{F}}=2$, $L_{\text{S0}}=L/5$,
$\lambda_{\text{F}}$ denotes the Fermi wave length.}
\label{fig3}
\end{figure}
In Fig.~\ref{fig3}, we show examples of spin--polarized conductances, 
plotted as a function of the inverse magnetic field, for a fixed Fermi
energy and zero Zeeman splitting. The total conductance $G_{\up \up}+
G_{\down \up}$ (for injection of up--spins) presents the usual step--like
behavior, whereas the spin--polarized ones have an irregular shape that
depends on the spin precession length.

\begin{figure}[t]
\begin{center}
\centering
\includegraphics[width=3.1in]{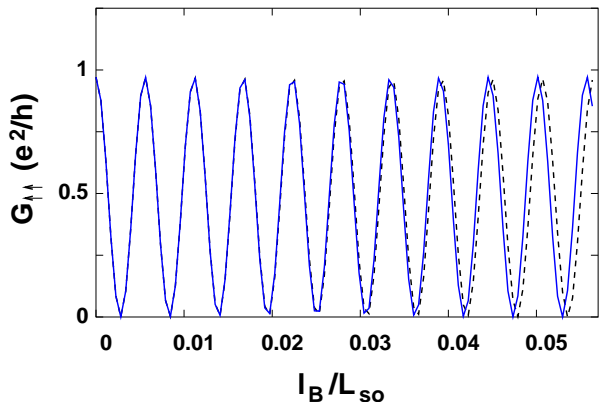}
\end{center}
\caption{(Color online) Conductance $G_{\up \up}$ of the spin--up edge
channel at filling factor one, computed both with transverse Rashba term
(solid line) and without it (dashed line) and plotted here as a function
of the Rashba SO coupling strength. The Zeeman coupling is set to zero.
The other parameters are: $E/\hbar \omega_{\text{c}}=1$, $L/
\lambda_{\text{F}}=40$, $W/\lambda_{\text{F}}=2$.}
\label{fig4}
\end{figure}
In Fig.~\ref{fig4}, we test the validity of the longitudinal SO
approximation. We plot $G_{\uparrow\uparrow}$ as a function of the 
SO coupling strength, computed with and without the longitudinal SO 
approximation. The two curves coincide perfectly for small Rashba
coupling and start differing slightly when the SO coupling becomes large.

\begin{figure}[b]
\begin{center}
\includegraphics[angle=0,width=0.4\textwidth]{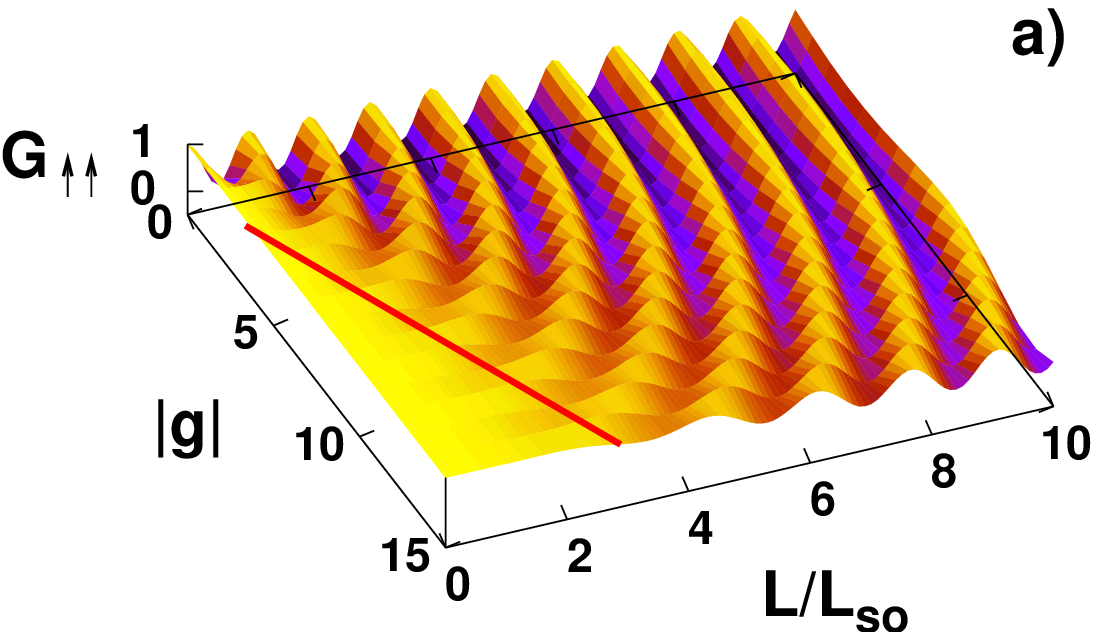}\\
\includegraphics[angle=0,width=0.4\textwidth]{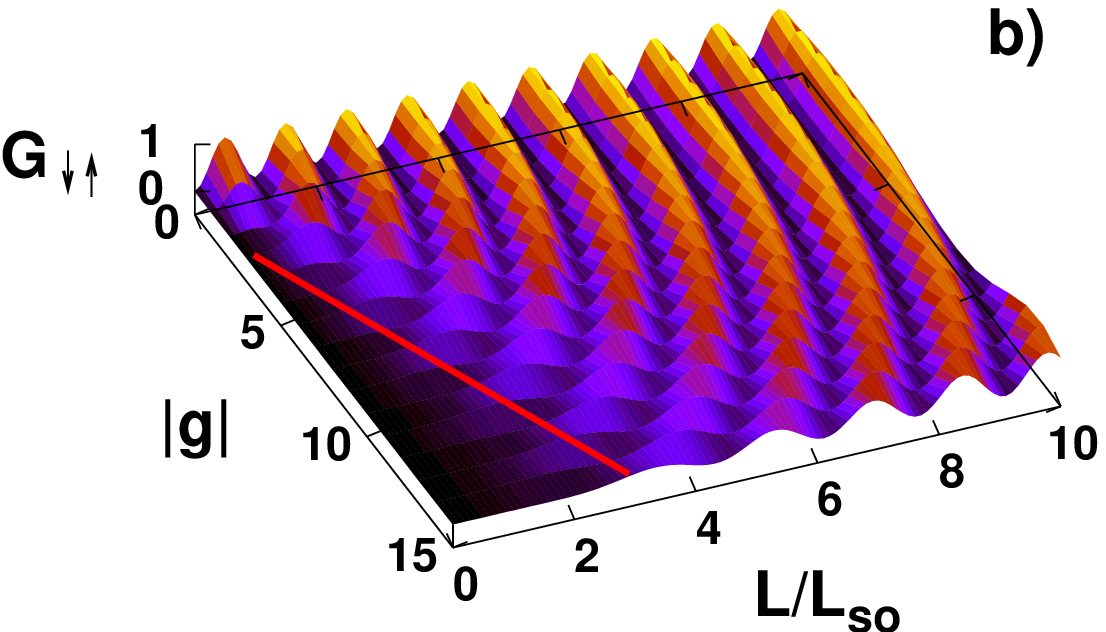}
\end{center}
\caption{(Color online) $G_{\up \up}$ (a) and $G_{\down \up}$ (b) plotted
as a function of SO coupling strength  and the g--factor $g$, at filling
factor 1. The red line represents the points for which $|\nu_{\rm R}/
\nu_{\rm Z}|=1$. (We computed $\nu_{\text{R}}$ assuming $v^{(0)}_n=
\omega_{\text{c}} l_B$, which is appropriate for a sharp edge potential.)
The other parameters are: $E_{\text{F}}/\hbar\omega_{\text{c}}=1$, 
$L/\lambda_{\text{F}}=40$, $W/\lambda_{\text{F}}=2$.}
\label{fig5}
\end{figure}
Now, we turn our attention to the interplay between the Rashba SO
coupling and the Zeeman splitting.\cite{meijer} In Fig.~\ref{fig5}, we
show spin--polarized conductances as a function of both $g$--factor and
SO  coupling strength (expressed in terms of the spin precession length). 
The effect of the Zeeman splitting is to induce a finite $z$--component in
the effective magnetic field $\mathbf{B}_{\text{eff}}$ around which the
spin precesses. When $g=0$, the effective field $\mathbf{B}_{\text{eff}}$
lies in the plane of the 2D electron system and, hence, it is orthogonal
to the $z$--axis and the current modulation is largest. In the opposite
limit, the Zeeman coupling tends to align the eigenspinors along the
$z$--direction, and spin-precession gets weaker. We conclude that the
presence of the Zeeman splitting has a negative effect on the spin
precession, but it does not disrupt it as long as $|\nu_{\rm R}/
\nu_{\rm Z}|>1$.

\section{Spin--dependent edge--channel interferometers}
\label{s5}

In this Section, we investigate the possibility to observe
spin--dependent interference effects between edge channels. This study is
motivated by a recent experimental realization of an electronic analog of
the optical Mach--Zehnder interferometer.\cite{heiblum} Spin--dependent
electron interferometry based on Rashba spin splitting has recently been
discussed, in {\em zero\/} magnetic field, in
Refs.~\onlinecite{absointerfer,uzapl04}.

\begin{figure}[b]
\begin{center}
\centering
\includegraphics[angle=0,width=0.4\textwidth]{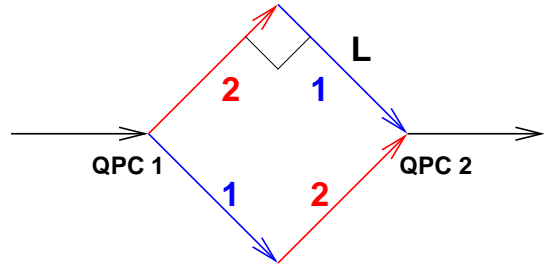}
\end{center}
\caption{(Color online)
Schematic view of the simplest interferometer setup. A right--moving edge
channel is split, via quantum point contact 1 (QPC1), and recombines at a
second quantum point contact (QPC2).}
\label{fig8}
\end{figure}
Edge channels are a very useful tool to construct electronic analogs of 
optical experiments, due to their chiral nature. Key elements of many
optical interferometers are beam splitters and wave guides. Both these
building blocks have been realized for electron waves in suitably
designed nanostructures.\cite{heiblum} The schematic setup of a possible
spin--dependent edge--channel interferometer is sketched in
Fig.~\ref{fig8}. A right--moving edge channel is split by QPC1 into two
different outgoing channels. These two states travel along a straight
segment of length $L$, perform an abrupt bend and, after another segment
of length $L$, interfere at QPC2. 

We consider the case of one propagating edge channel (filling factor one)
and the presence of both Rashba and Zeeman splittings. The question we
want to answer is whether a current modulation can be induced by varying
only the strength of Rashba SO coupling. Let us discuss, for the sake of
simplicity, the situation depicted in Fig.~\ref{fig8}. If the incident
spinor $\Psi_{\rm in}$ is split evenly at QPC1, the outgoing spinor at
QPC2 reads
\beq
\label{psi}
\Psi_{\rm out}=\frac{1}{\sqrt{2}} \left[
R_{\hat{n}_1} R_{\hat{n}_2} e^{+\pi i \Phi/\Phi_0}
+R_{\hat{n}_2} R_{\hat{n}_1} e^{-\pi i \Phi/\Phi_0} 
\right] \Psi_{\rm in}\;,
\eeq
where $\Phi$ is the magnetic flux encircled by the closed edge--channel
loop, $\Phi_0=h/e$ is the magnetic flux quantum, and $R_{\hat{n}_1,
\hat{n}_2}$ are rotation operators in  spin space which describe spin
precession. The precession axis is parallel to the direction of the
effective magnetic field, which depends on the directions of the electron
motion and the Zeeman field. In our simple setup, it can be directed
along $\hat{n}_1$ and $\hat{n}_2$, depending on the side of the
interferometer on which the electron is traveling. (See Fig.~\ref{fig8}.)
Explicitly, the rotation operator reads
\beq
\label{rot}
R_{\hat{n}_j}=
e^{-i \frac{\pi L}{L_{\text{so}}} \vec{\sigma} \cdot \hat{n}_j} \;,
\eeq 
with $j=1,2$, and $\hat{\sigma}$ denoting the vector of Pauli matrices. 
Let us consider the case $\Phi/\Phi_0=n$ (constructive interference due 
to the magnetic field). To obtain destructive interference in this
situation, we need that 
\beq
\label{cond}
R_{\hat{n}_1} R_{\hat{n}_2} +R_{\hat{n}_2} R_{\hat{n}_1} =0
\eeq
holds. If the sides of the interferometer are perpendicular and the
Zeeman term is negligible, the two directions $\hat{n}_1$ and $\hat{n}_2$
are orthogonal and, hence, $\sigma_{\hat{n}_1}\sigma_{\hat{n}_2}+
\sigma_{\hat{n}_2}\sigma_{\hat{n}_1}=0$. If this is the case, the
condition Eq.~(\ref{cond}) for destructive interference becomes $L=(n+1/2
)L_{\text{so}}$. A similar result was found in Ref.~\onlinecite{bercioux}
where the localization of electrons in quantum--coherent networks due to
Rashba SO coupling was discussed. We conclude that by varying the Rashba
coupling in our interferometer configuration realized with edge states,
it is possible to obtain a current modulation at a fixed magnetic field.
This setup would be a realization of the spin interferometer proposed in
Ref.~\onlinecite{absointerfer}, which is based on spin precession due to
the Rashba effect in a ring geometry. 

\section{Conclusions}
\label{conclusions}
We have obtained  analytical results, valid in the high--magnetic--field
regime, for quantum--Hall edge channels in the presence of Rashba
spin--orbit coupling and Zeeman splitting. Rashba spin precession is
expected to occur due to the relative shift of spin--polarized Landau
levels in guiding--center direction. Furthermore, we have presented
results on the effect of spin precession on edge--channel transport
obtained by a recursive Green's--function method. Finally, we have
proposed the realization of a spin--dependent interferometer based on
spin-precession of electrons in quantum--Hall edge channels. 

\begin{acknowledgments}
The authors would like to thank D.~Frustaglia for useful discussions
and helpful suggestions. This work has been supported by the IST NANOTCAD
project (EU contract IST-1999-10828), an EU Research Training Network
(RTN2-2001-00440), and the MacDiarmid Institute of Advanced Materials and
Nanotechnology (New Zealand).
\end{acknowledgments}


\end{document}